\documentclass[aps,prl,twocolumn,floatfix,showpacs,superscriptaddress]{revtex4}
\usepackage{graphics,graphicx}
\usepackage{amsfonts}
\usepackage{bm,bbm}
\usepackage{mathrsfs}
\usepackage{calc}
\usepackage{epsfig}
\usepackage{float}
\usepackage{color}
\usepackage{epstopdf}

\begin{document}

\title{Perturbation theory for quasi-energy (Floquet) solutions in the low frequency regime of the oscillating electric field}

\author{Hanna Martiskainen}
\affiliation{Physics Department,Technion - Israel Institute of Technology, Haifa 32000 Israel}
\author{Nimrod Moiseyev}
\affiliation{Physics Department,Technion - Israel Institute of Technology, Haifa 32000 Israel}
\affiliation{Schulich Faculty of Chemistry,Technion - Israel Institute of Technology, Haifa 32000 Israel}

\begin{abstract}
For a simple illustrative model Hamiltonian for Xenon in low frequency linearly polarized laser field we obtain a remarkable agreement between the zero-order energy as well
as amplitude and phase of the zero-order Floquet states and the exact eigenvalues and eigenfunctions of the Floquet operator. Here we use as a zero-order Hamiltonian the adiabatic Hamiltonian where time is used as an instantaneous parameter.  Moreover, for a variety of low laser frequencies, $\omega$,  the deviation of the zero-order solutions from the exact quasi-energy (QE) Floquet solutions approaches zero at the time the oscillating laser field is maximal. This remarkable result gives a further justification to the validity of the first step in the simple man model. It should be stressed that the numerical calculations of the exact QE (Floquet) solutions become extremely difficult when $\omega$ approaches zero and many Floquet channels are nested together and are coupled by the laser field.
This is the main motivation for the development of perturbation theory for QE (Floquet) solutions when the laser frequency is small, to avoid the need to represent the Floquet operator by a matrix when the Fourier functions are used as a basis set. A way to calculate the radius of convergence of the perturbational expansion of the Floquet solutions in $\omega$ is given.
\end{abstract}

\maketitle
\section{Motivation}
\subsection{The adiabatic approximation as it appears in the three-step model}

In 1993 Corkum\cite{corkum} and Kulander, Schafer and Krause\cite{kulander} showed that the three-step model (TSM) provides a classical interpretation of the high harmonic generation from atoms in strong laser fields without the need to solve numerically the time dependent Schr\"odinger equation (TDSE). This approach is known also as the simple man's theory. In 1994 Lewenstein, Balcou, Ivanov, L'Huillier and Corkum generalized the TSM to describe the interaction of arbitrary one-electron potentials with laser fields of arbitrary ellipticity and spectrum\cite{lewenstein}. This is, provided that the laser frequency is sufficiently low to insure that the Keldysh parameter is small enough\cite{lewenstein}.
In 2006 Santra and Gordon\cite{rs-ag-prl}, and Gordon,  K\"artner, Rohringer, and  Santra\cite{many-electron-3sm} generalized the TSM to atomic and molecular many-electron systems.

According to the TSM, in the first step an electron is excited to the continuum with no kinetic energy. This happens via tunneling through a potential barrier obtained when time is considered as an instantaneous parameter.
In the second step, the subsequent motion is governed classically by an oscillating electric field. In the third step, the electron returns to recombine with the parent nucleus. During this recombination process high order harmonics are emitted.

The first step in the TSM is an essential assumption in the derivation of this powerful and useful semiclassical approach. For this reason we focus here on the conditions under which the electron tunnels into the continuum through the adiabatic potential barrier. When these conditions are not met, the TSM is not applicable.
We wish to find out how well the zero-order solutions describe the amplitudes and the phases of the exact Floquet (quasi-energy) solutions. Particularly, we look for a formal justification of the successful applicability of the first step in the TSM, where the electron in assumed to tunnel through the adiabatic potential barrier near the peak of the laser field.

\subsection{On the validity of the adiabatic approximation and on the possibility to extend it to high frequency regime}

Within the adiabatic approximation time is treated as an instantaneous parameter.
The adiabatic hypothesis is based on a comparison between the (estimated) tunneling time with the period of the laser field. For sufficiently low laser frequencies, the tunneling time is smaller than the periodic time of the oscillating electric field, $T=2\pi/\omega$. The question we address ourselves in this paper is how can one extend this approach to higher laser frequencies. For sufficiently long laser pulses, the photoinduced dynamics can be described by the Floquet solutions which are eigenstates of the Floquet operator $\hat H^{ad}(t)-i\hbar \partial_t$. It is natural to apply perturbation theory where $\hat H^{ad}(t)$ is the zero-order Hamiltonian and $-i\hbar \partial_t$ is taken as a perturbation. By making a simple transformation to dimensionless time units $\tau=\omega t$, $\omega$ is obtained as the perturbational strength parameter.

In order to keep the adiabatic functions and energies as the leading dominant terms in the perturbation expansion of the Floquet eigenstates and eigenvalues, $\omega$ has to be a small parameter.
For large values of $\omega$, one should calculate the high-order terms in perturbation theory. Calculating the metastable photoionization states (so called resonances) using out going boundary conditions (as used by Gamow to calculate the tunneling decay rates for radioactive reactions), enables one to use the standard time-independent perturbation theory, as we will explain later.

The question is how far can we go with this approach? How large can $\omega$ be so that one can still calculate the Floquet resonances by using perturbation theory?
A perturbational approach, as we suggest here, is applicable when the perturbation series expansion converges. The method we use for calculating the radius of convergence avoids the need to calculate the high order terms in the perturbation series expansion of the eigenstates and eigenvalues of the Floquet operator. Even when the radius of convergence is zero, it might be possible to have an asymptotic expansion where the deviation of the partial sum of the terms from the exact solution is small, but at some point the error starts to increase. For example, the energy levels of atoms in a dc-field (known as the Stark effect) and the corresponding tunneling decay rates (inverse lifetimes) can be estimated from the perturbational corrections up to second order (where the field-free Hamiltonian is the zero-order Hamiltonian) for any value of the static field strength parameter. The radius of convergence in this case is zero.

As we will show here, the radius of convergence of the perturbation series expansion of the quasi-energies is finite and non-zero provided the expansion point is not a singularity. We will show that as the laser intensity decreases, the perturbation theory converges to the exact Floquet eigenstates and eigenvalues (which provide the photoionization lifetimes) for increasing laser frequencies, which serve as the perturbational strength parameter.

Our approach provides guide lines for future studies of the dynamics of high frequency strong lasers, based on perturbation theory where the adiabatic Hamiltonian, as it is used in the first step of the TSM, serves as the zero-order Hamiltonian of the perturbation expansion.

\section{Perturbation series expansion of the quasi-energy (QE) solutions}

Here we use the perturbation approach as first presented by Pont and his co-workers in Ref.\cite{PONT}.
The zero-order Hamiltonian is defined as
\begin{equation}
\label{H0}
\hat H^{(0)}(x,\tau)\equiv \hat {\cal H}(\tau)=\hat H_{ff}+\varepsilon_0\hat d\sin(\tau)=\hat H^{(0)}(x,\tau+2\pi).
\end{equation}
The perturbation operator is
\begin{equation}
 \hat H^{(1)}(x,\tau)=\hat H^{(1)}(x,\tau+2\pi)\equiv -i\hbar\frac{\partial}{\partial \tau}\,,
\end{equation}
and the perturbation strength parameter is $\omega$. The exact Floquet operator is given by
\begin{equation}
\label{FLOQUET}
\hat H_{Floquet}=\hat H^{(0)}(x,\tau)+\omega \hat H^{(1)}(x,\tau)
\end{equation}
We should re-emphasize that from now on we consider $\tau$ and $\omega$ to be two independent variables/parameters in spite that in the original Hamiltonian $\tau= t\omega$. Consequently, when the perturbation parameter, $\omega$, is set to zero then the zero-order Hamiltonian becomes the exact one.

The perturbation series expansion of the QEs is given by
\begin{equation}
{\cal E}_\alpha^{QE}={\cal E}_\alpha^{QE(0)}+\omega{\cal E}_\alpha^{QE(1)}+\omega^2{\cal E}_\alpha^{QE(2)}+...
\end{equation}

As it was proved in Ref\cite{PONT} the odd-order correction terms vanish. A simple explanation based on symmetry arguments is given below.
Complex absorbing boundary conditions are imposed on the solutions of the time dependent Schr\"odinger equation and on the Floquet solutions (within and without the framework of perturbation theory) such that the field-free bound states turn into metastable states (resonances). The photo-ionization decay rates are associated with the imaginary part of the complex QE eigenvalues.

\section{Zero-order QE solutions}
The zero-order Hamiltonian is the adiabatic Hamiltonian defined in Eq.\ref{H0}.
That is, $\hat {\cal H}(\tau)= \hat H^{(0)}(\tau)=\hat H^{(0)}(\tau+2\pi)$.
The $\tau$-periodic eigenvalues and the eigenfunctions of the zero-order Hamiltonian are given respectively by
\begin{eqnarray}
&& E^{(0)}_\alpha(\tau)=E^{(0)}_\alpha(\tau+2\pi) \\
&& \phi_\alpha^{(0)}(x,\tau)=\phi_\alpha^{(0)}(x,\tau+2\pi)\,.
\end{eqnarray}
We analytically continue $x$ to the complex plane $x=x'exp(i \theta)$ and therefore the zero-order eigenvalues are complex functions of $\tau$.
We might use only the resonance solutions in the construction of the zero-order Floquet solutions. We order the complex zero-order eigenvalues by the overlapping integral of the corresponding eigenfunctions with the field-free bound states. For example, $\alpha=1$ denotes the resonance solution which is associated with the ground state of the field-free Hamiltonian. As the field is turned on, the field-free ground state becomes a metastable (resonance) state. See Ref.\cite{JOH} for a summary of the time-independent perturbation theory and Ref.\cite{NM-PRC} for its extension to the complex scaled Hamiltonians and in Ref.\cite{NM-BOOK} for the use of time-independent propagation for time dependent Hamiltonian.

We use here the c-product such that
\begin{equation}
(\phi^{(0)}_{\alpha'}(\tau)|\phi^{(0)}_{\alpha}(\tau))\equiv \langle [\phi^{(0)}_{\alpha'}(\tau)]^*|\phi^{(0)}_{\alpha}(\tau)\rangle=\delta_{\alpha',\alpha}\,.
\end{equation}
Here we come to a delicate point in the perturbation derivation of the Floquet solutions. The zero-order quasi-energies are not $E_\alpha^{(0)}(\tau)$, but as usual in time-independent perturbation theory we should calculate expectation values by integrating over all independent variables of the full Hamiltonian (x and $\tau$ in our case).
Therefore,
\begin{eqnarray}
\label{ZERO-QE}
 &&{\cal E}^{QE(0)}_\alpha=\frac{1}{2\pi}\int_0^{2\pi}d\tau \int_{-\infty}^{+\infty} dx (\phi^{(0)}_{\alpha}(x,\tau)|\hat H^{(0)}|\phi^{(0)}_{\alpha}(x,\tau))\nonumber \\&&=
 \frac{1}{2\pi}\int_0^{2\pi}d\tau E_\alpha^{(0)}(\tau)\,.
 \end{eqnarray}
In Fig.~1 we show for a 1D model Hamiltonian, that the exact QE eigenvalues approach the value obtained by the zero-order calculations of ${\cal E}^{QE(0)}_{\alpha=1}$ as the laser frequency is reduced. Moreover, numerical fitting shows that the QEs are linearly proportional to $\omega^4$ and therefore the leading terms are the zero and the forth order perturbational terms.
To show that the exact quasi-energy periodic eigenfunctions of the Floquet operator are well described by the zero-order QE solutions for sufficiently low laser frequencies, we calculated the overlap between the zero-order QE functions and the exact QE(Floquet) solutions.
Our calculations show that the overlapping integral between the zero-order QE functions with the exact QE(Floquet) solutions of the 1D model Hamiltonian is very close to unity for a variety of laser frequencies. Larger overlap is obtained as the laser frequency is reduced. In Fig.~2 we present the deviation of the amplitudes of the zero-order QE (Floquet) functions from the exact QE functions as function of $\tau$ (the dimensionless time variable) for different values of the laser frequency $\omega$. As one can see the error is less than $10^{-3}$, ($0.1\%$), for the large frequency and it is reduced as $\omega$ is reduced.
In Fig.~3 we represent the deviation of the phases of the adiabatic solutions from the exact QE eigenfunction of the Floquet operator. As before, the deviation between the exact and the adiabatic solution is small and oscillates due to the oscillations of the laser field.
\emph{Our results presented in Figs.2-3 support the first step in the TSM \cite{corkum}-\cite{lewenstein}. It is remarkable indeed that for quite large values of the laser frequencies, the zero-order wavefunctions describe so well the exact Floquet solutions at the time the laser field gets its maximal amplitude.}\\

In order to minimize the deviation of the phase of the QE solutions obtained by the zero-order perturbation theory from the exact values, we multiply the eigenfunctions of the zero-order (adiabatic) Hamiltonian $\phi_\alpha^{(0)}(x,\tau)$ by a factor, $e^{if_\omega(\tau)}$, where $f_\omega(\tau)$ is \emph{real} and it is given by
\begin{equation}
 f_\omega(\tau)\\=\frac{1}{2\pi\omega}\int_0^{\tau}d\tau'Re[E_\alpha^{(0)}(\tau')-{\cal E}^{QE(0)}_\alpha]\,.
\end{equation}
The  solutions that include the phase factor above are defined as
\begin{equation}
\label{MODIFIED}
\widetilde{\phi}_\alpha^{(0)}(x,\tau)=e^{+if_\omega(t)}\phi_\alpha^{(0)}(x,\tau)\,.
\end{equation}
In Fig.~3 we show how by introducing the phase factor $f_\omega(\tau)$ the deviation of the phases of the modified adiabatic solutions from the exact values is reduced by several orders of magnitude and becomes almost $\omega$- independent (on this scale). In Fig.~4 we zoom on the results obtained when the adiabatic solutions are multiplied by the phase factor $\exp(i f_\omega(\tau))$. As one can see that on large scale the deviation of the phase of the modified functions from the exact value increases as $\omega$ is increased although on the scale of Fig.3 they seem to be $\omega$ independent. In a way, the fact that the contribution of high order terms to the power series expansion of the QE eigenvalues is reduced as $\omega$ is reduced reminds the increasing of the applicability of the semiclassical theory as $\hbar$ is reduced. It is important to notice that the phase factor as we defined here is different from the exponential \textbf{complex}  phase factor $\exp(i\int_0^\tau E_\alpha^{(0)}(\tau')d\tau'/(\hbar\omega))$, that was defined in the perturbation theory presented in Ref.\cite{PONT}.\\
When the zero-order wavefunctions describe so well the exact Floquet (QE) solutions, one might expect a non-zero radius of convergence. Note that this situation is very different from the standard perturbation theory where the zero-order Hamiltonian is the field-free Hamiltonian and the interaction with the field is taken as a perturbation. This is the well known Stark approach where the intensity of the laser is taken as the perturbation strength parameter (and not the frequency as we do) and the radius of convergence is zero.\\
\begin{figure}[htbp]
\includegraphics[width=1\columnwidth]{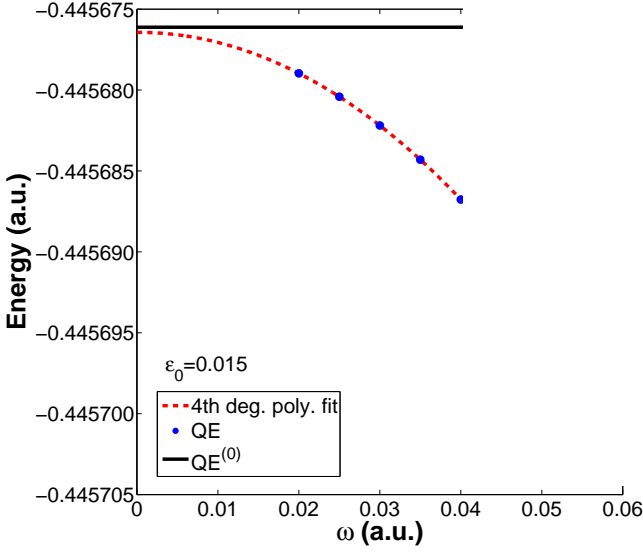}
\caption{The exact quasi-energies (QE) approach the zero-order QE (defined in Eq.\ref{ZERO-QE}) as the laser frequency is reduced. The calculations were carried for a 1D model Hamiltonian which mimics crudely a Xenon atom in linearly polarized laser field with laser field amplitude  $\varepsilon_0=0.015 \,a.u.$ which is $7.89\cdot10^{12} W/cm^2$ . Note that the first order derivative of the exact QE with respect to $\omega$ in our numerical fitting is equal to $-2.4\,10^{-6}$ in harmony with our proof that ${\cal E}^{QE(1)}_\alpha=0$ (see Eq.\ref{FIRST}).}
\end{figure}

\begin{figure}[htbp]
\includegraphics[width=1\columnwidth]{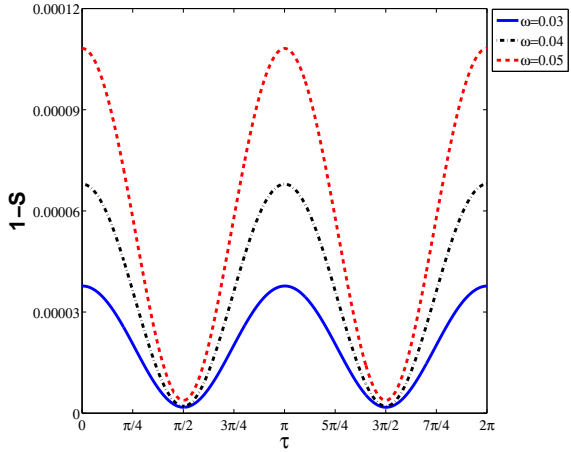}
\caption{The deviations from unity of the absolute value of the overlap integrals between the zero-order wavefunctions and the Floquet (QE) solutions which are associated with the field free ground state, (1-S), where $S=|\langle\Phi^{QE^{(0)}}(\tau)|\Phi^{QE}(\tau)\rangle|^{2}$. The calculations were carried for $\varepsilon_0=0.015 \,a.u.$ which is $7.89\cdot10^{12} W/cm^2$. As the frequency is reduced, the deviation of the amplitude of the zero-order QE solutions from the exact Floquet solutions is reduced. The minimal deviation from the exact QE solutions is obtained when the amplitude of the oscillating laser field reaches its maximal value. This result strongly supports the first step in the three-step model\cite{corkum}-\cite{lewenstein}. }
\end{figure}

\begin{figure}[htbp]
\includegraphics[width=1\columnwidth]{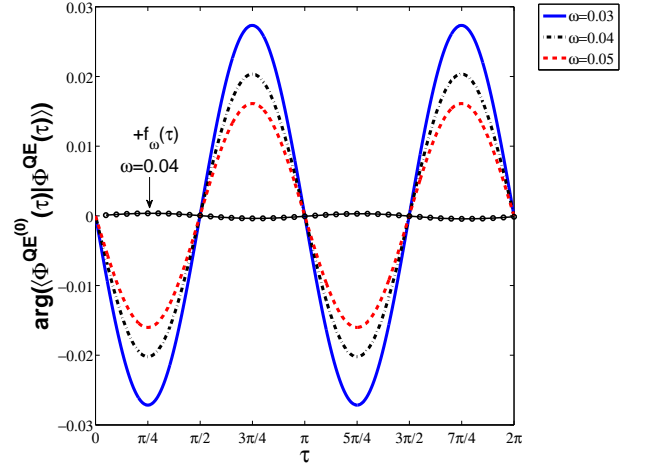}
\caption{The phases of the overlap integrals between the zero-order wavefunctions, ${\phi}_\alpha^{(0)}(x,\tau)$,  and the Floquet (QE) solutions which are associated with the field free ground state.  We show here that the deviations of phase of the zero-order solutions from the exact values are reduced by several orders of magnitude when we multiply the adiabatic  solutions by a time dependent phase factor (a zoom up is shown in Fig.4). It is a point of interest that the minimal (almost zero) phase deviation from the exact QE solutions is obtained when the amplitude of the oscillating laser field is zero or maximal.}
\end{figure}

\begin{figure}[htbp]
\includegraphics[width=1\columnwidth]{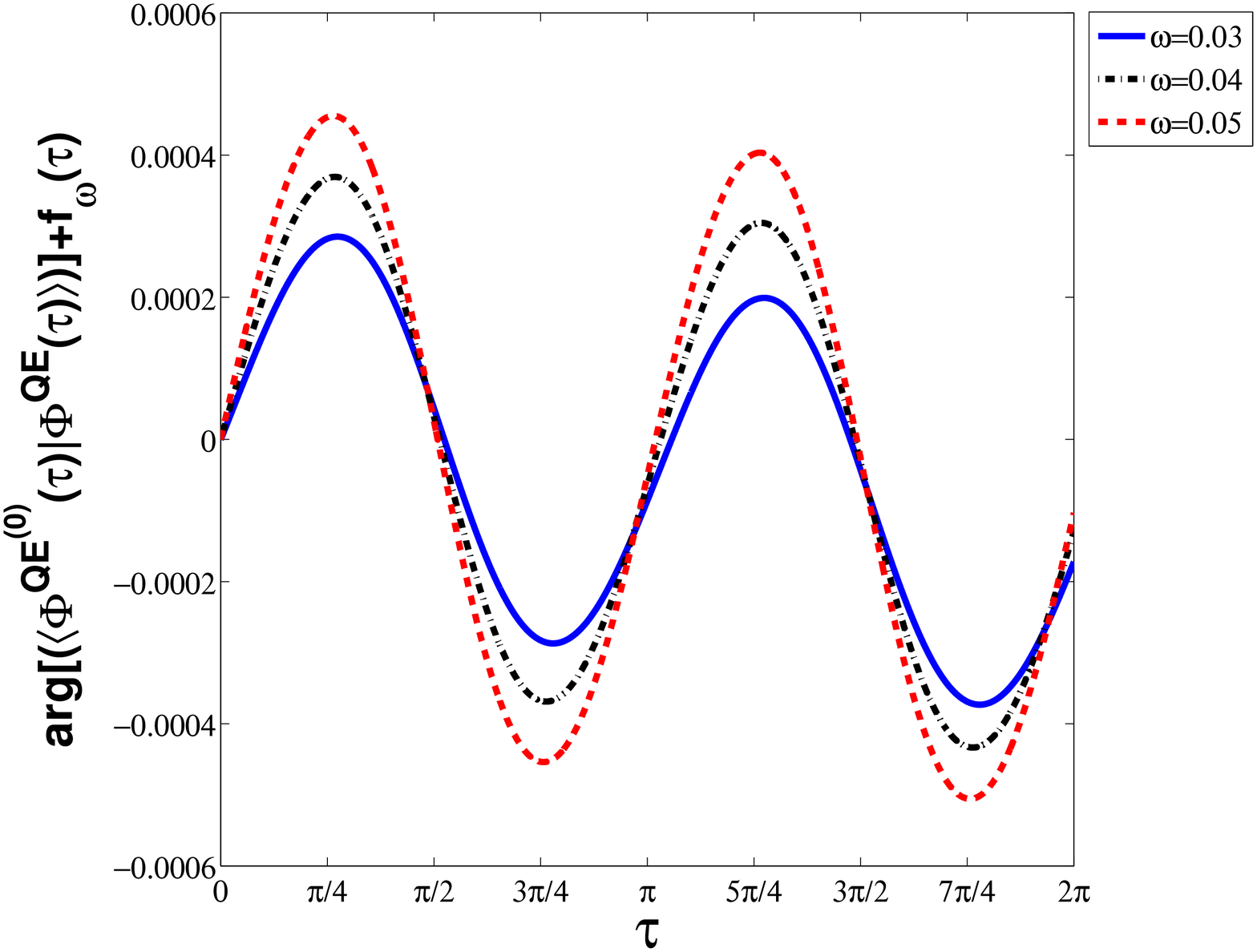}
\caption{The phases of the overlap integrals between the modified zero-order wavefunctions, $\widetilde{\phi}_\alpha^{(0)}(x,\tau)$ as defined in Eq.\ref{MODIFIED},  and the Floquet (QE) solutions which are associated with the field free ground state.
The phase of the overlap between the modified zero-order QE solutions and the exact Floquet solutions is reduced by several orders of magnitude in comparison to the results presented in Fig.3. }
\end{figure}

The zero-order solutions were calculated for a 1D effective model potential V(x), which crudely mimics a Xenon atom, and is given by
\begin{equation}
V(x)= -V_{0}exp(-a x^2)
\end{equation}
where $V_{0}=0.63$ and $a=0.1424$.\\
This potential has two bound states, with energies $\varepsilon_1=-0.4451 a.u.$ and $\varepsilon_2=-0.1400 a.u.$.
This model has been used before for calculating the HHG spectra of Xenon\cite{AVNER}. The atom interacts with a linearly polarized laser field $\varepsilon_0 xsin(\omega t)$.
Using the adiabatic approximation, time is used as an instantaneous parameter defined by $\tau=\omega t$. In Fig.5 we show how the potential varies with $\tau$.
\begin{figure}[htbp]
\includegraphics[scale=0.25]{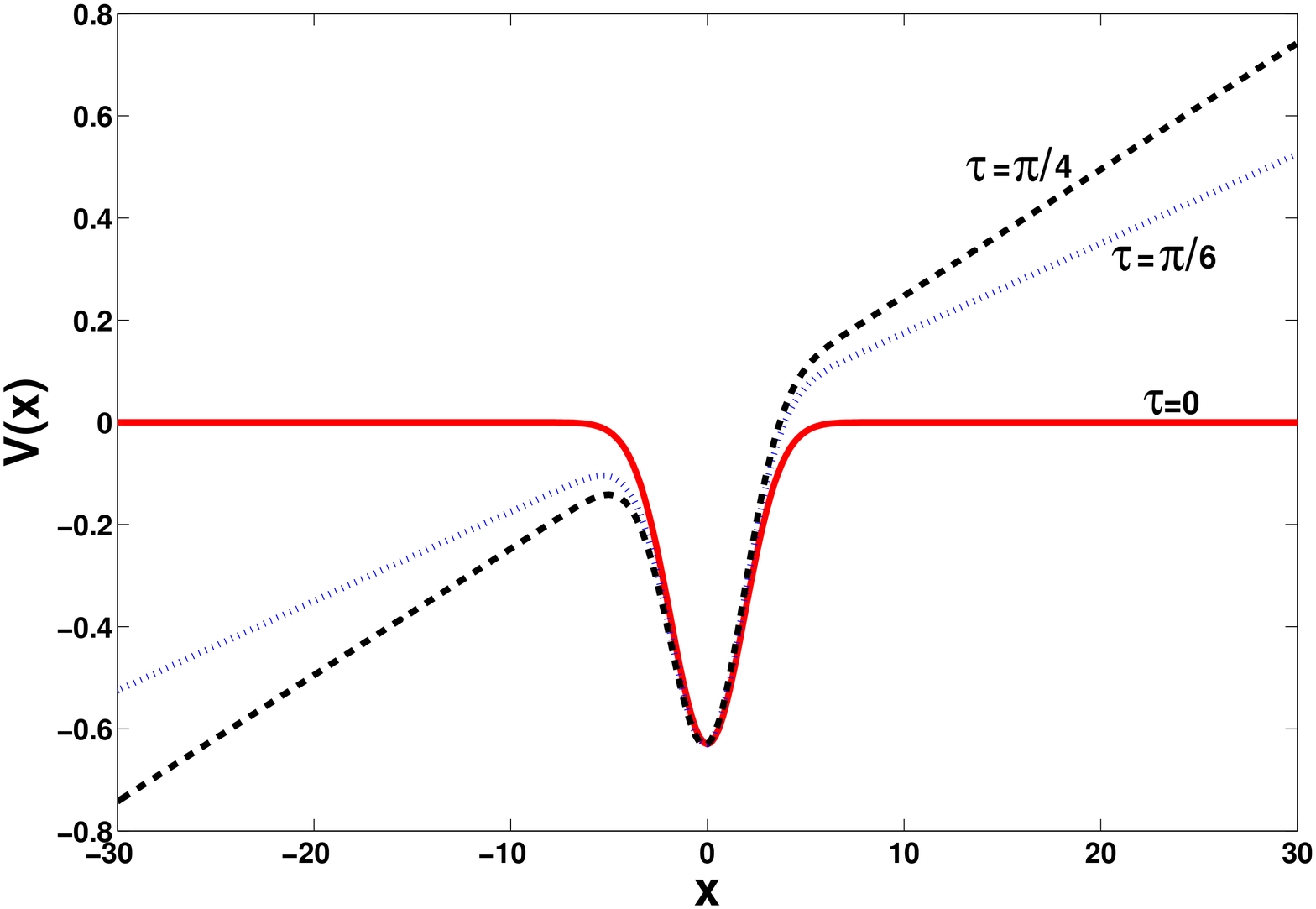}
\caption{The potential for a model Xenon atom represented by one dimensional Gaussian potential $V\left(x\right)=-0.63\exp{\left(-0.1424x^{2}\right)}$ which interacts with a static-like potential induced by the linearly polarized field $\varepsilon_{0}xsin(\tau)$. In this figure $\varepsilon_{0}=0.035$. The potential is displayed for different values of $\tau$.}
\end{figure}
For $\tau\ne 0$ the bound states of the field-free Hamiltonian turn into resonance states. We used the complex scaling transformation\cite{NM-BOOK}, i.e. $z=x\exp(i\theta)$, to impose outgoing boundary conditions.

 \section{Odd-order QE's}

 The first-order correction term to the perturbation series expansion of the QE (Floquet) energy is defined, as usual, as the expectation value of the perturbation using the zero-order functions. That is,
 \begin{eqnarray}
 &&{\cal E}^{QE(1)}_\alpha=\langle\langle \phi_\alpha^{(0)}|\hat H^{(1)}|\phi_\alpha^{(0)}\rangle\rangle_{x\tau} \nonumber \\ &&
 =\frac{1}{2\pi}\int_0^{2\pi}d\tau \int_{-\infty}^{+\infty} dx \phi^{(0)}_{\alpha}(x,\tau)\hat H^{(1)} \phi^{(0)}_{\alpha}(x,\tau)\nonumber \\&&= -\frac{1}{2\pi}\int_0^{2\pi}d\tau \frac{1}{2}(i\hbar \frac{\partial}{\partial \tau})\int_{-\infty}^{+\infty} dx \phi^{(0)}_{\alpha}(x,\tau)\phi^{(0)}_{\alpha}(x,\tau)\nonumber \\ &&=-i\hbar \frac{1}{4\pi}\int_0^{2\pi}d\tau \frac{\partial}{\partial \tau} 1=0\,.
 \end{eqnarray}

Our proof that
\begin{equation}
\label{FIRST}
\left[ \frac{\partial {\cal E}^{QE(exact)}_\alpha}{\partial \omega}\right]_{\omega=0}={\cal E}^{QE(1)}_\alpha=0
\end{equation}
has been confirmed in our numerical calculations presented in Fig.~1.

Based on same symmetry type arguments we reach the conclusions that all odd-order correction terms to the perturbational expansion of the quasi-energies are equal to zero since

\begin{eqnarray}
 &&{\cal E}^{QE(2n+1)}_\alpha=\langle\langle \phi_\alpha^{(2n)}|\hat H^{(1)}|\phi_\alpha^{(2n)}\rangle\rangle_{x\tau} \nonumber \\ &&
 =\frac{1}{2\pi}\int_0^{2\pi}d\tau \int_{-\infty}^{+\infty} dx \phi^{(2n)}_{\alpha}(x,\tau)\hat H^{(1)} \phi^{(2n)}_{\alpha}(x,\tau)\nonumber \\&&= -\frac{1}{2\pi}\int_0^{2\pi}d\tau \frac{1}{2}(i\hbar \frac{\partial}{\partial \tau})\int_{-\infty}^{+\infty} dx \phi^{(2n)}_{\alpha}(x,\tau)\phi^{(2n)}_{\alpha}(x,\tau)\nonumber \\ &&=-i\hbar \frac{1}{4\pi}\int_0^{2\pi}d\tau \frac{\partial}{\partial \tau} 1=0\,.
 \end{eqnarray}

These results are in a complete agreement with the proof given by Pont and his co-workers\cite{PONT} that the odd-terms in the Taylor series expansion of the \emph{exact} quasi-energies in $t\omega$ vanish.

\section{Radius of convergence}
For the association of radius of convergence of perturbation series expansion in a "small" parameter, $\lambda$, see Ref.\cite{TITCH}.
When the zero-order Hamiltonian and the perturbation commute, there is no branch point in the spectrum of $\hat H(\lambda)=\hat H^{(0)}+\lambda\hat H^{(1)}$ and the eigenvalues and eigenfunctions of $\hat H(\lambda)$ are analytical functions of $\lambda$. Consequently, perturbation theory converges for any value of $\lambda$. In our case, $\lambda$ is the perturbation strength parameter $\omega$ analytically continued to the complex regime. $\omega$, the physical laser frequency, gets only positive real values.
However, when $\hat H^{(0)}$ and $\hat H^{(1)}$ do not commute,
\begin{equation}
\left[\hat H^{(0)},\hat H^{(1)}\right]\ne 0\,,
\end{equation}
the perturbation expansion ${\cal E}_\alpha^{QE}={\cal E}_\alpha^{QE(0)}+\omega{\cal E}_\alpha^{QE(1)}+\omega^2{\cal E}_\alpha^{QE(2)}+...$
converges if and only if
\begin{equation}
\omega < |\lambda_{bp}|\,,
\end{equation}
where $\lambda_{bp}$ is the branch point closest to the origin, provided there is no singularity at $\omega=0$. That is,
\begin{equation}
\lambda_{bp}=|\lambda_{bp}|e^{i\gamma_{bp}}\,,
\end{equation}
for which the two eigenfunctions which have a dominant overlapping integral with the zero-order solution of interest ($\alpha=1$ for example), coalesce. In such a case, the exact QE solution is an analytical function of the complex $\lambda$ for any point which inside a circle with radius $|\lambda_{bp}|$. See for example Refs.\cite{NM-PRC},\cite{NM-SF} and references therein and Ref.\cite{NM-BOOK}. Let us add a technical explanation how the branch point can be calculated. When the branch point results from the coalescence of two QE solutions,
the difference between the values of two almost degenerate QE eigenvalues of the Floquet operator (as defined in Eq.\ref{FLOQUET}) is given by
\begin{equation}
{\Delta \cal E}_{\alpha,\alpha'}^{exact}(\omega)= a\sqrt{(\omega-\lambda_{bp})(\omega-\lambda^*_{bp})}
\end{equation}
for laser frequencies sufficiently close to $|\lambda_{bp}|$. $a$ is a complex prefactor. The complex parameters $a$ and $\lambda_{bp}$ can be computed from the high order terms in the perturbational series expansion in $\omega$ (for a given value of the maximum field amplitude). However, this procedure is quite complicated (but doable) as described in Refs.\cite{NM-PRC},\cite{NM-SF} and in Ref.\cite{NM-BOOK} on pages 331-333 (see also 235-237) and in the solution to Ex.~9.4. Here we present a new approach for the calculation of the branch point which determines the radius of convergence, based on the zero-order solutions which are easier to calculate than the exact QE solutions. In particular, it is hard to calculate the QE solutions for low laser frequencies. In the low frequency regime many Floquet channels are closely nested together and even a weak laser field couples them one to another such that the dimensions of the Floquet matrix required to get converged results become extremely large. Moreover, as it was shown in Appendix B in Ref.\cite{PONT}, the eigenvectors of the Floquet matrix as obtained when the periodic solutions are expanded in a Fourier basis functions  do not have a well defined limit as $\omega\to 0$. In other words, the radius of convergence of perturbation theory is zero when Fourier functions are used as a basis set. In this case the Floquet Hamiltonian is given by:
\begin{equation}
{\bf{\cal H}}_F={\bf{\cal H}}^{(0)}_F+\hbar\omega{\bf{\cal H}}^{(1)}_F
\end{equation}
where the Fourier matrix elements of the zero-order Hamiltonian and the perturbation are given by,
\begin{eqnarray}
&&[{\bf{\cal H}}^{(0)}_F]_{n',n} = \hat H_{ff}\delta_{n',n}+\frac{1}{2}\varepsilon_0\hat d\delta_{n',n\pm1} \nonumber \\
&&[{\bf{\cal H}}^{(1)}_F]_{n',n}= n\delta_{n',n}
\end{eqnarray}
where $n,n'=0,\pm 1,\pm2,...$

Note that the resonance solutions are obtained by imposing outgoing boundary conditions or by using one of the complex scaling transformations for which square integrable resonance solutions and rotating continuum are obtained. As mentioned above in Ref.\cite{PONT} it was shown that as $\omega\to 0$ the eigenvectors of ${\bf{\cal H}}_F$  (Eq. B1 in Ref.\cite{PONT}) do not have a well defined limit.
For this reason we developed a new approach for calculating the branch point (so called exceptional point) which determines the radius of convergence of the perturbation theory presented in this paper, without expanding the time periodic solutions in a Fourier basis.  The fact that indeed in our numerical calculations the phase of the Floquet solutions is obtained very accurately by the zero-order solutions for sufficiently small values of $\omega$, is an indication that there is no branch point at $\omega=0$ when the phase corrections to the zero-order solutions are taken as described above (see Figs.3-4).

\subsection{Calculation of the radius of convergence by diagonalizing the non-adiabatic time-dependent potential matrix}

The effective Hamiltonian, where the non-adiabatic couplings are taken into consideration, is given by a 2x2 matrix for the 2 level model Hamiltonian:
\begin{equation}
\label{HAM2by2}
{\bf H}_{eff}(\omega,\varepsilon_0,\tau)=\left(
  \begin{array}{cc}
    H_{11}(\varepsilon_0,\tau) & \omega H_{12}(\varepsilon_0,\tau) \\
   \omega H_{21}(\varepsilon_0,\tau) & H_{22}(\varepsilon_0,\tau) \\
  \end{array}
\right)
\end{equation}

where:
\begin{eqnarray}
\label{matrix-elements-HAM2by2}
  &&H_{11}(\varepsilon_0,\tau) =E_1^{ad}(\varepsilon_{0},\tau)\nonumber\\
  &&H_{12}(\varepsilon_0,\tau) =i \frac{\varepsilon_{0}\cos(\tau)}{E_2^{ad}(\varepsilon_{0},\tau)-E_1^{ad}(\varepsilon_{0},\tau)}\langle (\Psi^{ad}_1)^*|z|\Psi_2^{ad}\rangle\nonumber\\
  &&H_{21}(\varepsilon_0,\tau) =-H_{12}\nonumber\\
  &&H_{22}(\varepsilon_0,\tau) =E_2^{ad}(\varepsilon_{0},\tau)
\end{eqnarray}

As previously discussed, the radius of convergence of the perturbation series expansion in $\omega$ for a given set of parameters ($\varepsilon_0,\tau$) is obtained by replacing $\omega$ in Eq.\ref{HAM2by2} by a complex $\lambda$. We found $\lambda=\lambda_{BP}$ for which the spectrum of ${\bf H}_{eff}(\omega,\varepsilon_0,\tau)$ is degenerate.  This value is obtained when the discriminant of the second order polynomial of the eigenvalue solutions of ${\bf H}_{eff}(\omega,\varepsilon_0,\tau)$ vanishes. That is, a branch point in the spectrum of the full Hamiltonian, ${\bf H}_{eff}(\omega,\varepsilon_0,\tau)$, is obtained for

\begin{eqnarray}
\label{LAMBDA}
 \lambda_{BP}(\varepsilon_{0},\tau)= -\frac{(E_2^{ad}(\varepsilon_{0},\tau)-E_1^{ad}(\varepsilon_{0},\tau))^2}{2\varepsilon_{0}\cos(\tau)\langle (\Psi^{ad}_1(\varepsilon_{0},\tau))^*|z|\Psi_2^{ad}(\varepsilon_{0},\tau)\rangle}
\end{eqnarray}

By calculating the complex $\lambda_{BP}(\varepsilon_{0},\tau)$ for a given value of $\varepsilon_{0}$ as function of $\tau$  we obtained the value of the radius of convergence for the laser frequency  $\omega_{r.c.}(\varepsilon_{0})$:
\begin{equation}
 \omega_{r.c.}(\varepsilon_{0})\equiv Min\left|\lambda_{BP}(\varepsilon_{0},\tau)\right|_{\tau=\tau_0}.
  \end{equation}

Note that in the equation above $Min\left|\lambda_{BP}(\varepsilon_{0},\tau)\right|$ is the global minimum in the variation of
$\left|\lambda_{BP}(\varepsilon_{0},\tau)\right|$ as a function of $\tau$ (See Fig.~6).

\begin{figure}[htbp]
\includegraphics[width=1\columnwidth]{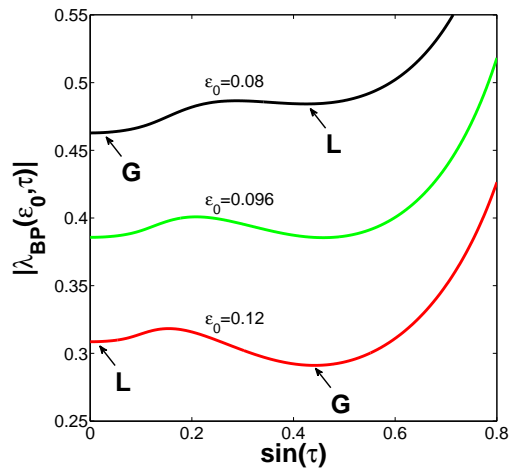}
\caption{The absolute value of the complex "frequency" at which a branch point is obtained in the spectrum of the time dependent Hamiltonian matrix, $|\lambda_{BP}(\varepsilon_{0},\tau)|$ as defined in Eq.\ref{LAMBDA}, is plotted as function of $\sin\tau$ for different values of the field amplitude, $\varepsilon_{0}$. "L" and "G" respectively denote the local and the global minima of $|\lambda_{BP}(\varepsilon_{0},\tau)|$. The global minimum "G" provides the radius of convergence for the laser frequency which is used as the perturbational strength parameter.}
\end{figure}

For a given value of the field amplitude, $\varepsilon_0$, the radius of convergence in perturbational expansion in $\omega$, which holds for any $0\le \tau\le 2\pi$, is given by

\begin{equation}
|\omega| < \omega_{r.c.}(\varepsilon_{0})
\end{equation}

In Fig.~6 we show that the upper limit of the laser frequency for which perturbation theory can possibly converge is \emph{reduced} as the field amplitude is increased.

\section{Concluding remarks}

In this work we have studied the application of perturbation theory for calculating the quasi-energy (Floquet) solutions, where in the zero-order Hamiltonian time is treated as an instantaneous parameter rather than a dynamical variable. Our results for a time periodic model Hamiltonian (which mimics the interaction of Xenon atom with a linearly polarized laser field) show that the overlapping integral (amplitude and phase) of the zero-order solutions with the exact Floquet solutions is remarkable for a quite large range of the laser frequencies. We have shown that the deviation of the zero-order solutions from the exact Floquet solutions is smallest when the oscillating laser field gets its maximal value. This result strongly supports the first step in the TSM, as described in Refs.\cite{corkum}-\cite{lewenstein}.

The radius of convergence is associated with a non-Hermitian degeneracy (a branch point) in the spectrum of the dressed-atom Hamiltonian. These branch points are very different in their nature from the physical branch points which are associated with exceptional points in the spectrum of the Floquet operator of time periodic Hamiltonian systems, since they are obtained as the laser frequency is analytically continued into the complex plane. Here we show how for a given laser intensity the range of laser frequencies for which the perturbation theory converges, can be calculated. As the laser field intensity is increased, the laser frequency for which the perturbation theory converges is reduced. These results agree with the expectations based on physical intuition. Yet, perturbation theory provides a rigorous method for improving the results obtained on the basis of the adiabatic hypothesis. In particular, it may be useful for laser frequencies which are high enough to question the validity of the adiabatic hypothesis and for going beyond the one electron model simulations and solving the full body problem, where the electronic correlation effects are taken into consideration.

\begin{acknowledgments}
ISF grant 298/11 and the I-Core: the Israeli Excellence Center "Circle of Light" are acknowledged for a partial support.
We acknowledge Dr. Ido Gilary from the Technion and Prof. Manfred Lein from the University of Hanover for their most fruitful discussions and comments.
\end{acknowledgments}

\end{document}